\documentclass[12pt,twoside,a4paper]{article}
\usepackage{amsmath,amssymb,latexsym,theorem,natbib,epsfig,color,subfigure}
\usepackage{multirow,graphicx,array,rotating,url}  
\usepackage{epstopdf,afterpage,wasysym}
\usepackage{verbatim}
\usepackage[normalem]{ulem}
\newfont{\msa}{msam10 scaled\magstep1}
\newfont{\ssmsa}{msam9}

\def\crps{\mathop{\hbox{\rm CRPS}}}
\def\crpss{\mathop{\hbox{\rm CRPSS}}}

\def\md{\mathrm{MD}}

\numberwithin{equation}{section}

\title{Calibration of wind speed ensemble forecasts for power generation}

\author{{\sc S\'andor Baran} and {\sc \'Agnes Baran} \vspace*{0.5cm}\\
         Faculty of Informatics, University of Debrecen\\
         Kassai \'ut 26, H-4028 Debrecen, Hungary
         }

        \date{}

\begin{document}
\pagestyle{myheadings}

\maketitle

\begin{abstract}
  In the last decades wind power became the second largest energy source in the EU covering 16\,\% of its electricity demand. However, due to its volatility, accurate short range wind power predictions are required for successful integration of wind energy into the electrical grid.
Accurate predictions of wind power require accurate hub height  wind speed forecasts, where the state-of-the-art method is the probabilistic approach based on ensemble forecasts obtained from multiple runs of numerical weather prediction models. Nonetheless, ensemble forecasts are often uncalibrated and might also be biased, thus require some form of post-processing to improve their predictive performance. We propose a novel flexible machine learning approach for calibrating wind speed ensemble forecasts, which results in a truncated normal predictive distribution. In a case study based on 100m wind speed forecasts produced by the operational ensemble prediction system of the Hungarian Meteorological Service, the forecast skill of this method is compared with the predictive performance of three different ensemble model output statistics approaches and the raw ensemble forecasts. We show that compared
with the raw ensemble, post-processing always improves the calibration of probabilistic and accuracy of point forecasts and from the four competing methods the novel machine learning based approach results in the best overall performance.

\end{abstract}

\section{Introduction}
\label{sec:sec1}

The increasing challenges caused by consequences of air pollution and emission of greenhouse gases highlight the importance of transition of energy production towards renewable energy sources. Besides the classical hydro power, in the last decades photovoltaic and wind energy fulfilled larger and larger part of energy demand. In 2020 the world set a new record by adding 93 GW of new wind turbines, so the total capacity of wind farms reached 744 GW covering 7\,\% of the global electricity demand \citep{wwea21}. In the EU (United Kingdom included), this proportion reached 16\,\%, and the (world) record is held by Denmark, where wind accounted for 48\,\% of the electricity consumed in 2020 \citep{we21}. However, wind energy poses serious challenges to traditional electricity markets, so accurate short range (between several minutes and a couple of days) prediction of wind power is of utmost importance for wind farm managers and electric grid operators.

Although the relation between wind speed and produced wind energy is nonlinear and might also be nonstationary, more reliable wind speed forecasts obviously result in more reliable predictions of produced electricity. Wind speed forecasts, similar to other meteorological variables, are based on numerical weather prediction (NWP) models describing atmospheric processes via systems of partial differential equations. The state of the art approach is to run an NWP model several times with different initial conditions which results in an ensemble of forecasts \citep{btb15}. Ensemble forecasts enable estimation of situation dependent probability distributions of future weather variables, which opens the door for probabilistic weather forecasting \citep{gr05}, where besides getting a point forecast the forecast uncertainty is also assessed. 

Recently all major weather centres operate their own ensemble prediction system (EPS), e.g. the 35-member Pr\'evision d'Ensemble ARPEGE\footnote{Action de Recherche Petite Echelle Grande Echelle (i.e. Research Project on Small and Large Scales)} (PEARP) EPS of M\'eteo France \citep{dljbac} or the 11-member Applications of Research to Operations at Mesoscale EPS \citep[AROME-EPS;][]{rvsz20} of the Hungarian Meteorological Service (HMS), whereas the largest ensemble size corresponds to the 51-member EPS of the European Centre for Medium-Range Weather Forecasts \citep{bpp98}. Nowadays ensemble weather forecasts are also popular inputs to probabilistic forecasts of renewable energy \citep{pm18}.

However, ensemble forecasts often appear to be uncalibrated and/or biased, this feature has been observed in several operational ensembles \citep[see e.g.][]{bhtp05}. A possible solution is the use of some form of statistical post-processing \citep{buizza18}, where nonparametric methods usually capture predictive distributions via estimating their quantiles \citep[see e.g.][]{fh07,brem19}, whereas parametric post-processing approaches provide full predictive distributions of the future weather quantities \citep[see e.g.][]{grwg05,rgbp05}. Recently machine learning based methods also gain more and more popularity \citep[see e.g.][]{rl18,tm20}; for a detailed overview of statistical calibration techniques we refer to \citet{wilks18} or \citet{vbde21}. 

Here we focus on a weather quantity important in energy production and investigate statistical post-processing of ensemble forecasts of wind speed measured at hub height (100m). In this context \citet{tsb09} proposes kernel dressing with Gaussian kernel left truncated at zero (TN; truncated normal), while \citet{mzbm13} considers forecasts based on inverse power curves and applies a censored normal predictive distribution. However, any post-processing method appropriate for wind speed can be applied, and we concentrate on the ensemble model output statistic \citep[EMOS;][]{grwg05} approach, where the predictive distribution is a single parametric probability law with parameters depending on the ensemble forecasts via appropriate link functions. To account for the non-negativity and right skew of wind speed, \citet{tg10} proposes a TN, \citet{bl15} a log-normal (LN), whereas \citet{bszsz21} a truncated generalized extreme value (TGEV) predictive distribution, and several methods for combining these probabilistic forecasts have also been developed \citep[see e.g.][]{lt13,bl16,bl18}. 

In the present paper we test the forecast skill of TN, LN and TGEV EMOS approaches on AROME-EPS forecasts of hub height wind speed. We also introduce a novel model with TN predictive distribution, where using the ideas of \citet{rl18} and  \citet{gzshf21}, location and scale parameters of the TN law are connected to the ensemble members via a multilayer perceptron neural network \citep[MLP;][]{dlbook}. Compared with the case of fixed link functions, this latter approach allows more flexibility in modelling and straightforward inclusion of new covariates as well. Note that TN, LN and TGEV EMOS approaches and some of their combinations have already been successfully applied for calibration of surface wind speed forecasts of the 11-member Aire Limit\'ee Adaptation dynamique D\'eveloppement International-Hungary Ensemble Prediction System of the HMS \citep{hkkr06}, see e.g. \citet{bhn14}.

The paper is organized as follows. In Section \ref{sec:sec2} the detailed description of the AROME-EPS is given, while in Section \ref{sec:sec3} the applied post-processing methods and considered verification tools are reviewed. The results of our case study is presented  in Section \ref{sec:sec4} followed by a concluding discussion in Section \ref{sec:sec5}.

\section{Data}
\label{sec:sec2}
The 11-member AROME-EPS 
of the HMS covers the Transcarpatian Basin with a horizontal resolution of 2.5 km \citep{rvsz20}. It consists of a control member and 10 ensemble members obtained from perturbed initial conditions. The dataset at hand contains ensemble forecasts of wind speed (m/s) at hub height (100m) together with the corresponding validation observations for three wind farms in the north-western part of Hungary (\'Acs, J\'anossomorja and P\'apakov\'acsi) for the period 7 May 2020 to 28 March 2021. All forecasts are initialized at 0000 UTC with a temporal resolution of 15 minutes and maximal forecast horizon of 48h resulting in a total of 192 forecast lead times.

\section{Post-processing methods and verification tools}
\label{sec:sec3}

Non-homogeneous regression or EMOS is one of the most popular parametric post-processing approaches, probably due to its computational efficiency and excellent performance for a wide range of weather variables. EMOS models for different weather quantities differ in the parametric family specifying the predictive distribution; however, most of the existing EMOS models are implemented in the {\tt ensembleMOS} package of {\tt R} \citep{emos}.

In the following sections let \ $f_1,f_2, \ldots , f_{11}$ \ denote the 11-member AROME-EPS hub height wind speed forecast for a given location, time and lead time, where \ $f_1=f_{\text{CTRL}}$ \ is the control forecast, while \ $f_2,f_3, \ldots ,f_{11}$ \ correspond to the $10$ statistically indistinguishable (and thus exchangeable) ensemble members \  $f_{\text{ENS},1},f_{\text{ENS},2}, \ldots ,f_{\text{ENS},10}$  \ generated using random perturbations.  Further, let \ $\overline f$ \ denote the ensemble mean, \ $\overline f_{\text{ENS}}$ \ the mean of the $10$ exchangeable members, and denote by \ $S^2$ \ and \ $\md$ \ the ensemble variance and ensemble mean absolute difference, respectively, defined as
\begin{equation*}
S^2:=\frac 1{10}\sum_{k=1}^{11}\big(f_k-\overline f\big)^2 \qquad \text{and} \qquad \md:=\frac 1{11^2}\sum_{k=1}^{11}\sum_{\ell=1}^{11}\big|f_k-f_{\ell}\big|.
  \end{equation*}

\subsection{Truncated normal EMOS model}
  \label{subs3.1}
Let \ $\mathcal{N}_0\big(\mu,\sigma^2\big)$ \ denote the TN distribution  with location \ $\mu$, \ scale \ $\sigma > 0$, \ and lower truncation at $0$, having probability density function (PDF)
\begin{equation*}
    g(x|\mu,\sigma) := 
    \frac1{\sigma}\varphi\big((x-\mu)/\sigma\big) / \Phi(\mu/\sigma), \qquad \text{if \ $x\geq 0$,}
  \end{equation*}
  and \  $g(x|\mu,\sigma):=0$, \ otherwise, where \ $\varphi$ \ is the PDF, while \ $\Phi$ \ denotes the cumulative distribution function (CDF) of a standard normal distribution. The proposed TN EMOS predictive distribution for hub height wind speed based on the AROME-EPS ensemble forecast is
  \begin{equation}
    \label{tn_emos}
    \mathcal{N}_0\big(a_0 + a^2_{\text{CTRL}}f_{\text{CTRL}}+a^2_{\text{ENS}}\overline f_{\text{ENS}}, b^2_0 + b^2_1\md\big),
  \end{equation}
where \ $a_0,a_{\text{CTRL}},a_{\text{ENS}},b_0,b_1\in {\mathbb R}$.  \  The same model is applied by \citet{hemri14} to model square root of 10m wind speed, and the suggested method  is a slight modification of the TN EMOS approach of \citet{tg10}, where the square of the scale parameter is an affine function of the ensemble variance, that is \ $\sigma^2=b^2_0+b^2_1S^2$. \  Exploratory tests
with the dataset at hand show that neither modelling the square root of the data, nor linking location to the ensemble variance result in better forecast skill than the use of \eqref{tn_emos}.

\subsection{Log-normal EMOS model}
\label{subs3.2}
As an alternative to the TN EMOS approach, we consider the EMOS model of \citet{bl15}, where the mean \ $m$ \ and variance \ $v$ \ of the LN predictive distribution are affine functions of the ensemble members and the ensemble variance, respectively, that is
\begin{equation}
   \label{ln_emos}
  m=\alpha _0 + \alpha^2_{\text{CTRL}}f_{\text{CTRL}}+\alpha^2_{\text{ENS}}\overline f_{\text{ENS}} \qquad \text{and} \qquad  v=\beta^2_0 + \beta^2_1S^2,
\end{equation}
where $\alpha_0,\alpha_{\text{CTRL}},\alpha_{\text{ENS}},\beta_0,\beta_1\in {\mathbb R}$. \ The heavier upper tail of the LN distribution allows a better fit to high wind speed values.

\subsection{Truncated generalized extreme value EMOS model}
\label{subs3.3}

Another possible solution to address reliability of probabilistic forecasts for high wind speed is the use of the GEV EMOS approach proposed by \citet{lt13}. The GEV distribution \ $\mathcal{GEV}(\mu,\sigma,\xi)$ \ with location \ $\mu$, \ scale \ $\sigma >0$ \ and shape \ $\xi$ \ is defined by CDF
\begin{equation}
\label{eq:gevCDF}
G(x|\mu,\sigma,\xi) := \begin{cases}
\exp\Big(-\big[1+\xi(\frac{x-\mu}{\sigma})\big]^{-1/\xi}\Big), & \quad \text{if \ $\xi \neq 0$;}\\
\exp\Big(-\exp\big(-\frac{x-\mu}{\sigma}\big)\Big),  & \quad \text{if \ $\xi = 0$},
\end{cases}
\end{equation}
for \ $1+\xi(\frac{x-\mu}{\sigma}) > 0$ \ and \ $G(x|\mu,\sigma,\xi) := 0$, \ otherwise. However, as demonstrated by \citet{lt13} and \citet{bl15}, the GEV EMOS model might assign positive predicted probability to negative wind speed. 
To correct this deficiency, \citet{bszsz21} propose to truncate the GEV distribution from below at zero and consider a TGEV predictive distribution  \ $\mathcal{TGEV}(\mu,\sigma,\xi)$ \ with location \ $\mu$, \ scale \ $\sigma >0$ \ and shape \ $\xi$ \  defined by CDF
\begin{equation}
    \label{eq:tgevCDF}
    G_0(x|\mu,\sigma,\xi)=\begin{cases}
\frac{G(x|\mu,\sigma,\xi)-G(0|\mu,\sigma,\xi)}{1-G(0|\mu,\sigma,\xi)}, & \text{if \ $G(0|\mu,\sigma,\xi) < 1$};\\
1,  & \text{if \ $G(0|\mu,\sigma,\xi) = 1$},
\end{cases} 
\end{equation}
for \ $x\geq 0$, \ and  \ $G(x|\mu,\sigma,\xi) := 0$, \ otherwise.

For the 11-member AROME-EPS, location and scale parameters of the TGEV EMOS model are
\begin{equation}
\label{eq:tgevlocscale}
\mu = \gamma_0 + \gamma_{\text{CTRL}}f_{\text{CTRL}} + \gamma_{\text{ENS}}\overline f_{\text{ENS}}\qquad \text{and} \qquad \sigma = \sigma^2_0 + \sigma^2_1\overline{f},
\end{equation}
with \ $\gamma_0,\gamma_{\text{CTRL}},\gamma_{\text{ENS}},\sigma_0,\sigma_1\in {\mathbb R}$, \  while the shape parameter \ $\xi$ \ does not depend on the ensemble members.
In order to ensure a finite mean and positive skewness, the shape is kept in the interval \ $]-0.278,1/3[$.

\subsection{Parameter estimation}
\label{subs3.4}
Parameter estimation in the TN, LN and TGEV EMOS models described in Sections \ref{subs3.1} -- \ref{subs3.3} is based on the optimum score principle of \citet{gr07}. The estimates are obtained as minimizers of the mean value of a proper scoring rule over an appropriate training dataset. Here we consider one of the most popular proper scores in atmospheric sciences, namely the continuous ranked probability score \citep[CRPS;][Section 9.5.1]{w19}. Given a (predictive) CDF \ $F$ \ and a real value (observation) \ $x$, \ the CRPS is defined as 
\begin{equation}
    \label{eq:CRPSdef}
\crps(F,x) := \int_{-\infty}^{\infty}\Big[F(y)-{\mathbb I}_{\{y\geq x\}}\Big]^2{\mathrm d}y ={\mathsf E}|X-x|-\frac 12
{\mathsf E}|X-X'|,
\end{equation}
where ${\mathbb I}_H$ \ denotes the indicator function of a set \ $H$, \ while \ $X$ \ and \ $X'$ \ are independent random variables distributed according to \ $F$ \ and having a finite first moment.  CRPS is a negatively oriented score, that is the smaller the better, and the right hand side of \eqref{eq:CRPSdef} implies that it can be expressed in the same units as the observation. Note that the CRPS for TN, LN and TGEV distributions can be expressed in closed form (see \citet{tg10}, \citet{bl15} and \citet{bszsz21}, respectively), which allows an efficient optimization procedure.

A crucial issue in statistical calibration is the selection of training data. Here the different forecast horizons are treated separately and we use rolling training periods, which is a standard approach in EMOS modelling. In this training scheme parameters for a given lead time are estimated with the help of corresponding forecast--observation pairs from the preceding \ $n$ \ calendar days. Further, both regional (or global) and local EMOS models are investigated. In the regional approach all data from the training period are considered together, providing  a single set of EMOS parameters for all three wind farms. In contrast, local estimation results in different parameter estimates for different wind farms by using only data of the given location. In general, local models outperform their regional counterparts \citep[see e.g.][]{tg10}, provided the training period is long enough to avoid numerical stability issues \citep{lb17}.

\subsection{Machine learning based approach to wind speed modelling}
\label{subs3.5}

As mentioned in the Introduction, based on works of \citet{rl18}
and \citet{gzshf21}, we applied a machine learning approach to estimate
the parameters of the predictive distribution in a TN model.  In this
case, instead of looking for the parameters \
$a_0,a_{\text{CTRL}},a_{\text{ENS}},b_0,b_1$ \ in \eqref{tn_emos},
location and scale are estimated directly, without assuming that they
depend on the ensemble in a prescribed way. Practically this means,
that some features derived from the ensemble e.g. the control member, or the ensemble standard deviation) are used as inputs of a
multilayer perceptron (MLP), while the trained network provides a
two-dimensional vector corresponding to the location and scale
parameters. Similar to the previous models, the network is trained
by minimizing the mean CRPS over the training data.

In an MLP some hidden layers connect the input layer and the output one, the number of layers and the number of neurons in the different hidden layers are tuning parameters of the network. Starting from the first hidden layer, each neuron of the given layer computes a weighted sum of the values provided by the neurons in the previous layer, adds a bias and via a so-called transfer function applies a transformation to the result.  

In the present work we train an MLP with one hidden layer containing 25 neurons, the applied transfer functions are the exponential linear unit \cite[ELU; see e.g.][]{gzshf21} function in the hidden layer, and the linear function in the output layer. After some experiments, in the final training  we decided to use the control forecast, the mean of the exchangeable ensemble members and the standard deviation of the 11 members as input features of the network. Based on \citet{gzshf21}, to ensure the positivity of the location and scale parameters, their estimates are given by \ $\exp ({\theta _1})$ \ and \ $\exp({\theta _2})$, \ where \ $\theta _1$ \ and \ $\theta _2$ \ are the values provided by the two neurons of the output layer.

By the training of a network the number of the training samples is
always a critical point: a relatively small training set can easily
result in  overfitting, which means a weak performance on the test
set. In order to avoid this problem we apply a regional estimation,
moreover, we do not handle the different lead times separately; for a
given training period we train only two networks, one for the 0--24h
forecasts, another for the 24--48h forecasts. We made a trial to take
into account the lead time in the training by extending the features
with a fourth one, containing the ranks of the lead times; however, this modification did not improve the predictive performance of the network. The lack of significance of the forecast horizon might be explained by the diurnal cycle in the ensemble standard deviation, which indicates a direct relation between forecast uncertainty and lead time.

\subsection{Verification tools}
\label{subs3.6}
As argued by \citet{gbr07}, the aim of probabilistic forecasting is to maximize the sharpness of the predictive distribution subject to calibration. The former refers to the concentration of the predictive distribution, whereas the latter means a statistical consistency between the validating observation and the corresponding predictive distribution. These goals can be addressed simultaneously using proper scoring rules quantifying the forecast skill by numerical values assigned to pairs of probabilistic forecasts and validating observations. In the case study of Section \ref{sec:sec4}, for a given lead time competing forecasts in terms of probability distribution are compared with the help of the mean CRPS over all forecast cases in the verification data. The improvement in terms of CRPS of a probabilistic forecast \ $F$ \ with respect to a reference forecast \ $F_{\text{ref}}$ \ can be assessed with the continuous ranked probability skill score \citep[CRPSS; see e.g.][]{gr07} defined as
$$\crpss := 1 - \frac{\overline\crps_F}{\overline\crps_{F_{\text{ref}}}},$$
where \ $\overline\crps_F$ \ and \ $\overline\crps_{F_{\text{ref}}}$ \ denote the mean score values corresponding to forecasts \ $F$ \ and \ $F_{\text{ref}}$, \ respectively. Here larger values indicate better forecast skill compared to the reference method.

Calibration and sharpness can also be quantified by the coverage and average width of the \ $(1-\alpha)100\,\%, \ \alpha \in ]0,1[$, \ central prediction interval, where calibration is defined as the proportion of validating observations located between the lower and upper \ $\alpha/2$ \ quantiles of the predictive distribution. For a well calibrated forecast, this value should be around \ $(1-\alpha)100\,\%$, \  and in order to provide a fair comparison with the 11-member AROME-EPS, \ $\alpha$ \ should be chosen to match the nominal  coverage  of $83.33\,\%$ ($10/12 \times 100\,\%$) of the raw ensemble.  

Simple graphical tools for assessing calibration of probabilistic forecasts are the verification rank histogram of ensemble predictions and its continuous counterpart, the probability integral transform (PIT) histogram. Verification rank is defined as the rank of the verifying observation with respect to the corresponding ensemble forecast \citep[Section 9.7.1]{w19}, whereas PIT is the value of the predictive CDF evaluated at the observation \citep[Section 9.5.4]{w19}. For a properly calibrated ensemble, all ranks should be equally likely, while calibrated predictive distributions result in standard uniform PIT values.

Finally, the accuracy of point forecasts, such as median and mean, is quantified with the help of mean absolute errors (MAEs) and root mean squared errors (RMSEs), respectively.

\section{Results}
\label{sec:sec4}

\begin{figure}[t]
  \centering
\epsfig{file=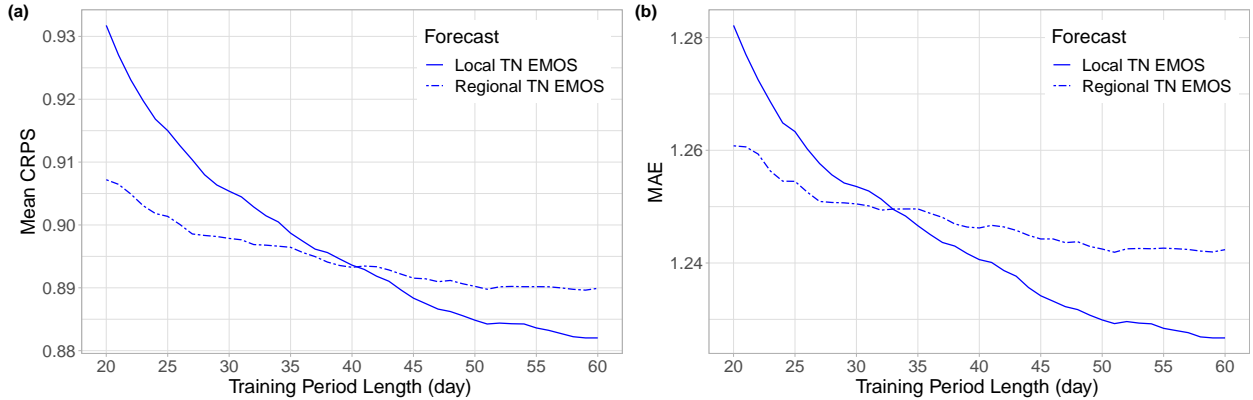, width=\textwidth}
\caption{Mean CRPS of probabilistic (a) and MAE of median (b) forecasts for local and regional TN EMOS models as functions of training-period length.}
\label{fig:train}
\end{figure}

We start our analysis by determining the appropriate training-period length for our post-processing approaches. We consider a fixed verification period from 8 July 2020 to 28 March 2021 (264 calendar days) and compare the forecast skill of both local and regional TN EMOS models estimated using \ $20, 21, \ldots , 60$ \ day rolling training-periods. Figure \ref{fig:train} shows the mean CRPS taken over all forecast cases and lead times and the MAE of median forecasts as functions of the training-period length. Both plots clearly demonstrate that for longer training periods the local TN EMOS is more skillful than the regional one. CRPS and MAE of the latter stabilize after day 51, while the corresponding scores of the local TN EMOS also seem to level off there. Hence, for TN EMOS modelling a 51-day training-period seems to be a reasonable choice, and the same training period length is applied for LN and GEV EMOS models as well. A detailed data analysis confirmed that this length is also appropriate for the machine learning approach of Section \ref{subs3.5} (TN MLP), this choice of training data leaves a total of 273 calendar days (period 29 June 2020 -- 28 March 2021) for model verification. Further, as in general, local versions of the tested EMOS approaches slightly outperform the regional ones, in what follows only the scores of the local models are reported.

\begin{figure}[t]
  \centering
\epsfig{file=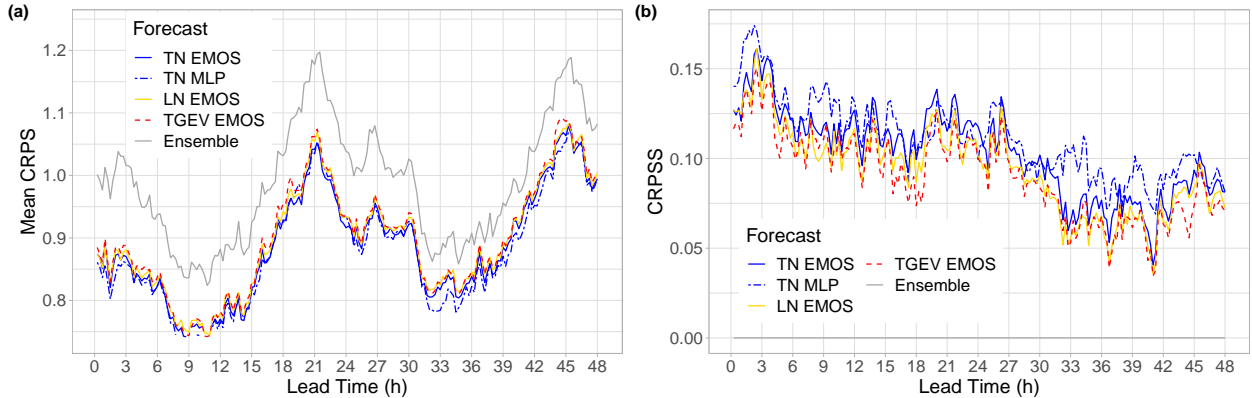, width=\textwidth}
\caption{Mean CRPS of post-processed and raw ensemble forecasts of wind speed (a) and CRPSS with respect to the raw ensemble (b) as functions of lead time.}
\label{fig:crps_lead}
\end{figure}

Figure \ref{fig:crps_lead}a shows the mean CRPS of post-processed and raw ensemble forecasts as functions of the lead time, whereas in Figure \ref{fig:crps_lead}b the corresponding CRPSS values with respect to the raw ensemble are plotted. In general, all post-processing approaches outperform the raw-ensemble for all lead times, but the advantage of post-processing decreases with the increase of the forecast horizon.  The best overall CRPSS taken over all lead times and forecast cases belongs to the TN MLP model ($0.111$), followed by the local TN EMOS method ($0.103$); however, there are certain forecast horizons (especially around 20h and 23h), where the latter exhibits slightly better predictive performance. For the TGEV and LN EMOS approaches these overall CRPSS values are $0.091$ and $0.095$, respectively. 

\begin{figure}[t]
  \centering
\epsfig{file=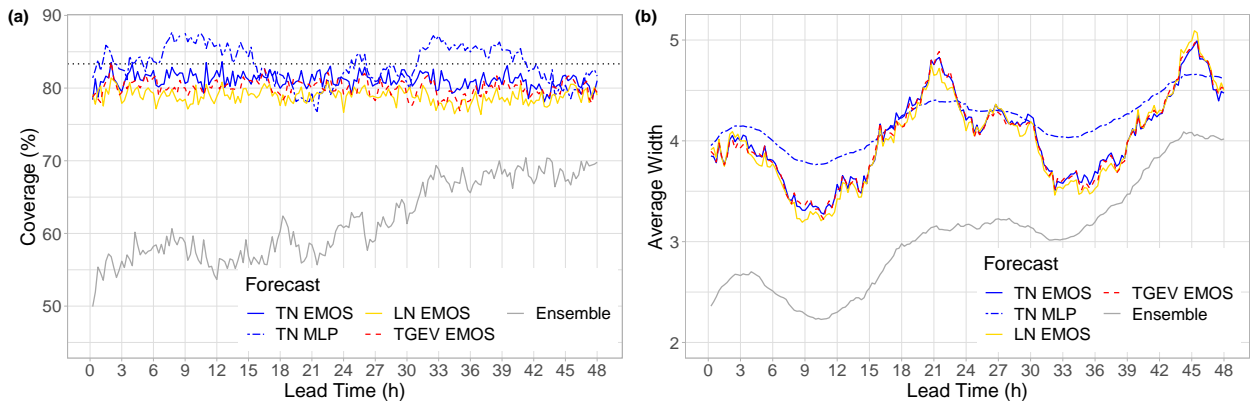, width=\textwidth}
\caption{Coverage (a) and average width (b) of the nominal 83.33\,\% central prediction intervals of post-processed and raw forecasts as functions of lead time.}
\label{fig:arome_cov_lead}
\end{figure}

\begin{figure}[t]
  \centering
\epsfig{file=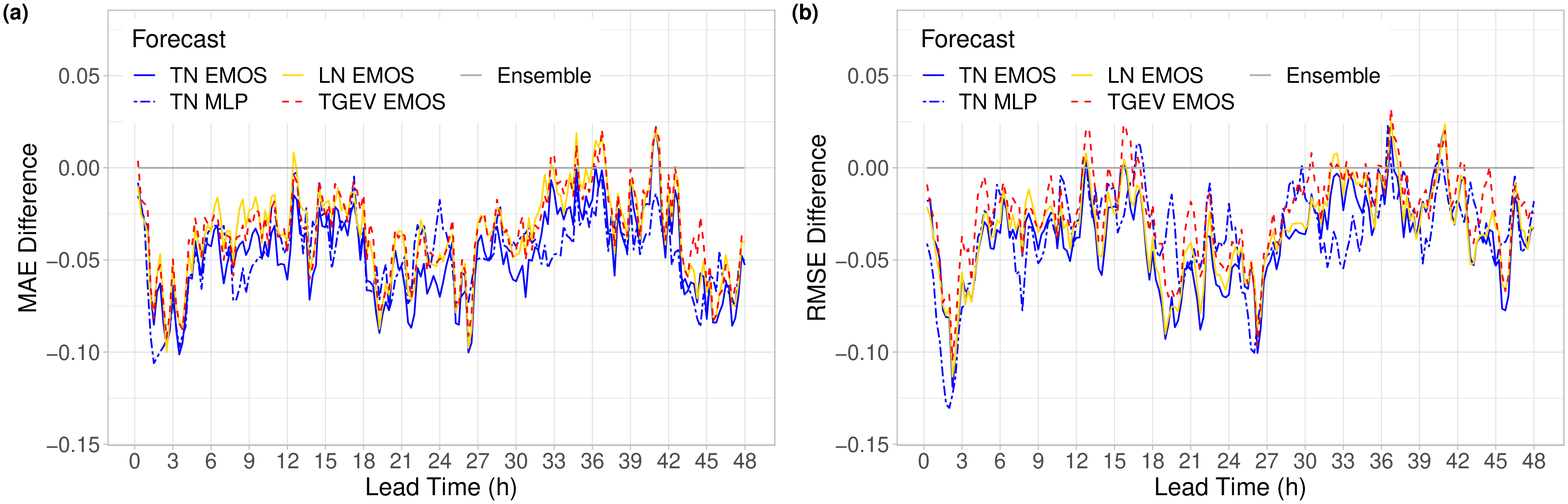, width=\textwidth}
\caption{Difference in MAE of the median forecasts (a) and in RMSE of the mean forecasts (b) from the raw ensemble as functions of lead time.}
\label{fig:arome_mae_lead}
\end{figure}

The improved calibration of post-processed forecasts can also be observed on Figure \ref{fig:arome_cov_lead}a showing the coverage of the nominal 83.33\,\% central prediction intervals for different lead times. The coverage of the AROME-EPS  ranges from 50\,\% to 70\,\% and in general, increases with the increase of the lead time, whereas all post-processed forecasts for all lead times result in coverage values that are rather close to the nominal level.  In particular, there is no visible systematic difference in the coverage values of the three investigated EMOS models, whereas the TN MLP approach seems to exhibit some kind of diurnal cycle. However, as depicted in Figure \ref{fig:arome_cov_lead}b, the cost of the better calibration should be paid in the deterioration of the sharpness. The raw ensemble produces far the narrowest central predictive intervals, there is no difference in sharpness between the competing EMOS models, whereas the diurnal cycle in sharpness of the TN MLP is completely in line with the corresponding coverage. Note that similar diurnal cycles can be observed in the ensemble standard deviation and ensemble mean difference as well.

\begin{figure}[t!]
  \centering
\epsfig{file=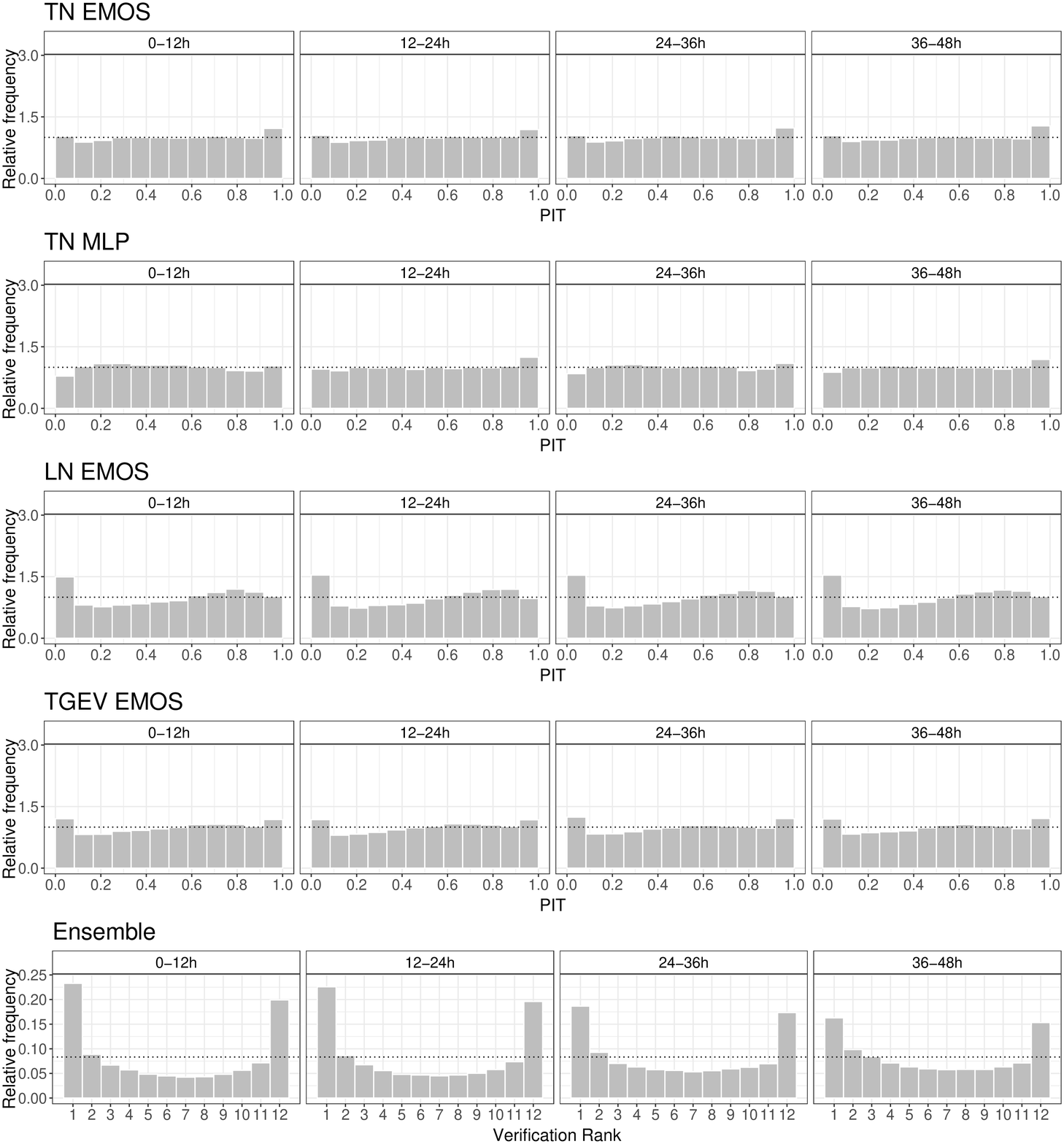, width=\textwidth}
\caption{PIT histograms of post-processed and verification rank histograms of raw ensemble forecasts of wind speed for the lead times 0-12h, 12-24h, 24-36h and 36-48h.}
\label{fig:arome_pit}
\end{figure}

While statistical post-processing substantially improves the calibration of probabilistic forecasts, it does not really effect the accuracy of point predictions. In Figure \ref{fig:arome_mae_lead}a the difference in MAE of the median forecasts of the various calibration methods from the MAE of the raw ensemble are plotted as functions of the lead time. Similar to the mean CRPS, models with TN predictive distribution show the best performance for all lead times; however, even the largest difference in MAE is less than $0.1$\,m/s. The same behaviour can be observed in Figure \ref{fig:arome_mae_lead}b displaying the difference in RMSE of the mean forecasts. This can indicate that the raw AROME-EPS forecasts are already unbiased and indeed, the mean biases of the ensemble mean and median  taken over all forecast cases of the whole available period 8 May 2020 to 28 March 2021 and all lead times are just $0.136$\,m/s and $0.122$\,m/s, respectively, while the overall MAE equals $1.285$\,m/s and the overall RMSE is $1.669$\,m/s.

Finally, Figure \ref{fig:arome_pit} shows the verification rank histograms of raw and PIT histograms of post-processed forecasts for four  different lead time intervals. The U-shaped verification rank histograms clearly indicate the underdispersive character of the raw ensemble; however, the dispersion improves with the forecast lead time. This behaviour is completely in line with the increasing coverage and high sharpness of the raw forecasts (see Figure \ref{fig:arome_cov_lead}). Further, the depicted rank histograms are rather symmetric, which is consistent with the small overall MAE and RMSE and illustrates the lack of bias in the raw ensemble. All post-processing approaches substantially improve calibration; models based on TN predictive distributions result in almost flat PIT histograms, whereas the histograms of TGEV and LN EMOS approaches indicate slight biases. Kolmogorov–Smirnov (KS) test rejects the uniformity of the PIT for all models; however, based on the values of the KS test statistic, one can provide a clear ranking of the methods. PIT values of  the TN MLP approach fit best the uniform distribution, followed by the TN, TGEV and LN EMOS models, which order nicely reflects the shapes of the corresponding histograms of Figure \ref{fig:arome_pit}.

\begin{figure}[t]
  \centering
\epsfig{file=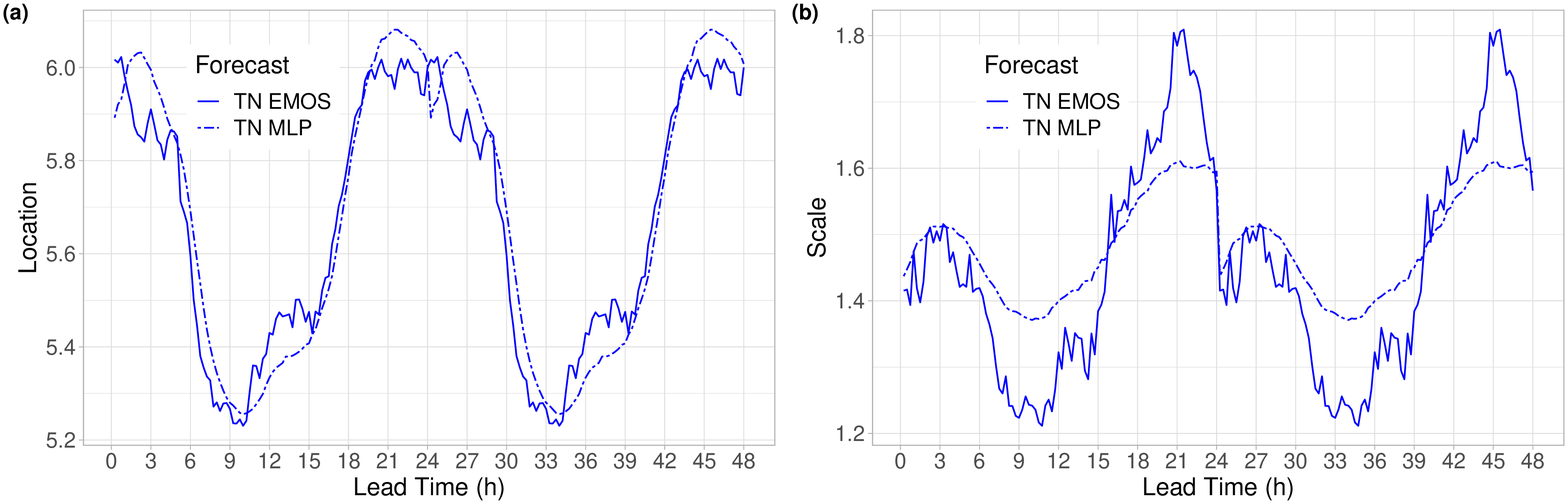, width=\textwidth}
\caption{Mean of the location (a) and scale (b) of the truncated normal predictive distributions of TN EMOS and TN MLP models as functions of lead time.}
\label{fig:pars}
\end{figure}

Based on the above results, one can conclude that models with TN predictive distributions provide the best forecast skill, and the machine learning based TN MLP approach outperforms the TN EMOS model. Hence, one might be interested in the dissimilarities of the corresponding predictive distributions. According to Figure \ref{fig:pars}a, there is no fundamental difference in the location, and the station-wise time series plots of this parameter also provide matching curves (not shown). Thus the linear model of the location given in \eqref{tn_emos} seems to be optimal. A completely different picture can be observed in Figure \ref{fig:pars}b, showing the mean of the scales of the TN predictive distributions as function of lead time. The diurnal cycle for TN MLP is far less pronounced than for the TN EMOS, and the corresponding time series (not shown) exhibit completely different behaviour, too. Hence, the superior performance of the TN MLP approach is due to the more general modelling of the scale of the TN predictive distribution.

\section{Conclusions}
\label{sec:sec5}
We investigate post-processing of ensemble forecasts of 100m wind speed, as this variable is of crucial interest in wind energy production. 
Three  different EMOS models based on truncated normal, log-normal and truncated generalized extreme value distributions are considered and we also propose a novel method where the probabilistic forecasts are obtained in the form of a truncated normal predictive distribution with parameters linked to the ensemble via a multilayer perceptron neural network. The forecasts skill of the competing calibration methods is tested on the 11-member AROME-EPS hub height wind speed ensemble forecasts of the HMS for three wind farms in Hungary and verified against observations provided by the wind farm operators. Only short term predictions are considered with forecast horizons ranging up to 48h with a temporal resolution of 15 minutes. Using the raw ensemble as reference, we compare the mean CRPS of probabilistic, MAE of median and RMSE of mean forecasts and the coverage of central prediction intervals corresponding to the nominal 83.33\,\% coverage. We also study the shapes of the PIT histograms of the calibrated forecasts for different lead times and compare with the corresponding verification rank histograms of the raw ensemble. Based on our case study we can conclude that compared with the raw ensemble, post-processing always improves the calibration of probabilistic and accuracy of point forecasts. From the four competing methods the novel machine learning based TN MLP approach exhibits the best overall performance; moreover, in contrast to the investigated EMOS models, it provides a single universal model for several forecast horizons. The superior performance of the TN MLP model is explained by its ability to represent more complex nonlinear relations between the ensemble forecasts and the parameters of the TN predictive distribution and our results are consistent with the findings of \citet{rl18} and \citet{gzshf21}.

The present work highlights several directions of potential future research. From the one hand, one might consider the machine learning approach to parameter estimation in the case of other predictive distribution families such as the LN and TGEV investigated here. From the other hand, a neural network allows a very flexible choice of input features, providing a simple and straightforward opportunity of involving predictions of other weather variables in wind speed modelling.

\bigskip
\noindent
{\bf Acknowledgments.} \  S\'andor Baran was supported by the National Research, Development and Innovation Office under Grant No. NN125679. The authors thank Gabriella Sz\'epsz\'o and Mih\'aly Sz\H ucs from the HMS for providing the AROME-EPS data.

\end{document}